\documentclass[]{raa}
\usepackage{graphicx,times}
\usepackage{natbib}
\usepackage{amssymb,amsmath}
\bibpunct{(}{)}{;}{a}{}{,}

\usepackage[a4paper=true,dvipdfm=true,pagebackref=true]{hyperref}
\hypersetup{pdftitle = The title of my PDF, pdfauthor = My name, pdfsubject= The subject, pdfkeywords = keyword1 keyword2 keyword3}
\hypersetup{colorlinks = true, linkcolor = green, anchorcolor = red, citecolor = blue, filecolor = red, pagecolor = red, urlcolor = red}

\begin{document}

   \title{The Deconvolution of Lunar Brightness Temperature based on Maximum Entropy Method using Chang'E-2 Microwave Data
$^*$
\footnotetext{\small $*$ Supported by the National Natural Science Foundation of China.}
}

 \volnopage{ {\bf 2012} Vol.\ {\bf X} No. {\bf XX}, 000--000}
   \setcounter{page}{1}

   \author{Shu-Guo Xing\inst{}, Yan Su\inst{}, Jian-Qing Feng\inst{}, Chun-Lai Li\inst{}
   }

   \institute{National Astronomical Observatories,Chinese Academy of Sciences,Beijing 100012,
China; {\it suyan@nao.cas.cn}\\
\vs \no
}

\abstract{A passive and multi-channel microwave sounder onboard Chang'E-2 orbiter has successfully performed microwave observation of the lunar surface and subsurface structure. Compared with Chang'E-1 orbiter, Chang'E-2 orbiter obtained more accurate and comprehensive microwave brightness temperature data which is helpful for further research. Since there is a close relationship between microwave brightness temperature data and some related properties of the lunar regolith, such as the thickness, temperature and dielectric constant, so precise and high resolution brightness temperature is necessary for such research. However, through the detection mechanism of the microwave sounder, the brightness temperature data acquired from the microwave sounder is weighted by the antenna radiation pattern, so the data is the convolution of the antenna radiation pattern and the lunar brightness temperature. In order to obtain the real lunar brightness temperature, a deconvolution method is needed. The aim of this paper is to solve the problem in performing deconvolution of the lunar brightness temperature. In this study, we introduce the maximum entropy method(MEM) to process the brightness temperature data and achieve excellent results. The paper mainly includes the following aspects: firstly, we introduce the principle of the MEM, secondly, through a series of simulations, the MEM has been verified an efficient deconvolution method, thirdly, the MEM is used to process the Chang'E-2 microwave data and the results are significant.
\keywords{Space vehicles---instruments: microwave sounder---Moon: brightness temperature---method:maximum entropy method
}
}

   \authorrunning{S.-G. Xing et al. }            
   \titlerunning{The Deconvolution of Lunar Brightness Temperature}  
   \maketitle

\clearpage
%
\section{Introduction}           
\label{sect:intro}

Chang'E-2 orbiter was launched on October 1, 2010. The microwave sounder is one of the main payloads on both Chang'E-1 and Chang'E-2 orbiter, which is passive and has four channels(3GHz, 7.8GHz, 19.35GHz, 37GHz)(\citealt{1jiang2010china}). During on-orbit observation, both of the two microwave sounders obtained a lot of microwave radiation brightness temperature data. Many scholars have begun to carry out extensive researches using these microwave brightness temperature data({\citealt{2fa2010primary},\citealt{3wang2010lunar}). Benefiting from the lower orbit altitude, the spatial resolution of the microwave sounder on Chang'E-2 orbiter is nearly twice as good as Chang'E-1. In addition, Chang'E-2 measurements covered the lunar surface more completely and got more tracks of swath data than Chang'E-1. Therefore, more available data can be used to research radiation characteristics of lunar surface and subsurface structure.
In this paper, we choose the 37GHz microwave data, which better reflect the lunar surface microwave radiation. Moreover, the Laser Altimeter and the Charge Coupled Device(CCD) camera are also the main payloads on the Chang'E-2 orbiter, thus the Digital Elevation Model (DEM) and CCD images are direct response to lunar surface structure. So we use the DEM and CCD images to verify the results of the deconvolution.

It is known that in the detection mechanism of the microwave sounder, antenna brightness temperature has been weighted by antenna radiation pattern. It is the convolution of the antenna radiation pattern and the lunar brightness temperature. The result of this impact is shown in Figure \ref{fig1}.  Because of the impact of the antenna pattern, points close together on the original map (Fig. 1a) can't be distinguished in the observation map (Fig.1c).
Actually, when using the brightness temperature data to research, we suppose the lunar brightness temperature is approximately equal to the antenna brightness temperature, but this will cause the loss of details and a decrease in spatial resolution(\citealt{9feng2013data}). The paper introduces a deconvolution method named Maximum Entropy Method(MEM) to eliminate the influence of the antenna pattern. In the field of image reconstruction of radio astronomy, MEM is a valuable tool that has many advantages in dealing with both the extended sources and the point sources(\citealt{4starck2002deconvolution}).
\begin{figure}[htb]
   \centering
   \includegraphics[width=14.0cm, angle=-1]{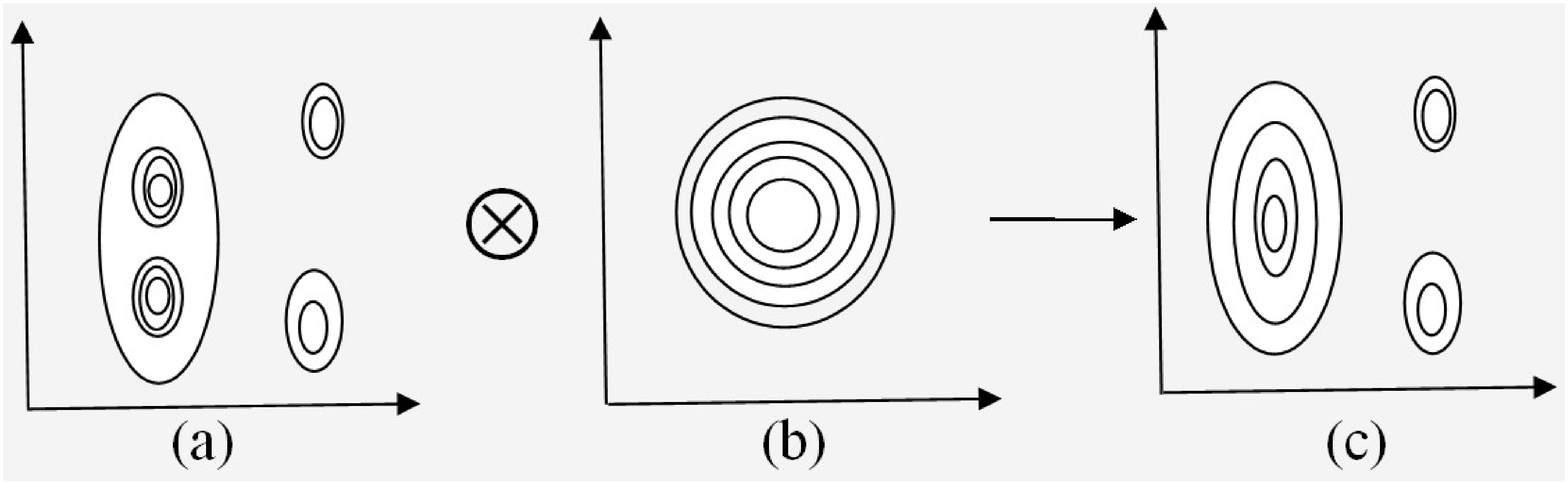}
   \caption{The diagram of the impact of the antenna pattern. (a) the original map (b) the antenna pattern (c) the observation map(the convolution of (a) and (b)). }
   \label{fig1}
\end{figure}

The paper is organized as follows. In Section 2, a brief description on the principle of MEM is given. In Section 3, a series of simulations are carried out to verity the validity of MEM. In this section, we introduce two indices, Mean Squared Error(MSE) and Peak Signal to Noise Ratio(PSNR), to evaluate the quality of the deconvolution. Through the simulations, we come to a conclusion that the MEM is an efficient deconvolution method. In Section 4, the MEM is used to process the Chang'E-2 brightness temperature data. A new brightness distribution map is shown and compared with the original map, and we also choose several local regions for discussions of the results. Finally, conclusion of this work is stated in Section 6.

\section{The principle of MEM}

In the field of image restoration, Frieden first introduced the concept of image entropy in 1972(\citealt{5frieden1972restoring}) and gave the expression, $S =  - \sum {x(i,j)\ln x(i,j)}$, where $x$ defined the gray value of the image at the point $i$ and $j$. With the introduction of image entropy, a new constraint condition has been applied in the image restoration. Based on the different background and the limitations of the application of the MEM, Frieden, Gull and Bryan proposed different MEMs named by the authors name respectively(\citealt{5frieden1972restoring};\citealt{6gull1978image};\citealt{7skilling1984maximum}). The MEM used in the paper is based on method proposed by N.L.Bonavito in 1993 ({\citealt{8bonavito1993maximum}}).

Here, a model used in the paper is shown as below.

~~~

$\left\{ \begin{array}{l}
subject ~to ~y(i,j) = x(i,j) \otimes PSF\\
max ~S =  - \sum {x(i,j) \cdot \ln x(i,j)} \\
find ~x(i,j)
\end{array} \right.$

~~~

In the above model, $x$ means the original map, $y$ means the observation map, $PSF$ means the antenna pattern, and the model indicates that the best solution of the deconvolution need obey the constraint of the maximum entropy.

With the lagrange multiply method, we get the maximum entropy distribution,

\begin{equation}\label{equ1}
    x(i,j) = \exp ( - \sum\limits_{m,n = 1,1}^{i,j} {\lambda mn}  \cdot PSF)/Z({\lambda _{11}}, \cdot  \cdot  \cdot ,\lambda ij)
\end{equation}

\begin{equation}\label{equ2}
    Z({\lambda _{11}}, \cdot  \cdot  \cdot ,\lambda ij) = \sum\limits_{1,1}^{i,j} {\exp ( - \sum\limits_{m,n = 1,1}^{i,j} {\lambda mn \cdot PSF} )}
\end{equation}
where $Z$ means the partition function, $\lambda$ means lagrange multipliers. Also, a parameter is defined as

\begin{equation}\label{equ3}
    G = x(i,j) \otimes PSF - y(i,j)
\end{equation}

Here we use a more storage-efficient method of successive approximation. Our recursion relation is defined as

\begin{equation}\label{equ4}
    \lambda (New) = \lambda (old) - \ln (\frac{y}{{y - G}})
\end{equation}

Steps are as follows: the first step, we set all $\lambda=0$; the second step, with the $\lambda=0$, we get an original $x$ from Equation(\ref{equ1}); the third step, we put $x$, $PSF$ and $y$ in the Equation(\ref{equ3}), and we calculate $G$; the fourth step, having the value of $G$, we get the new $\lambda$ from Equation(\ref{equ4}). Then based on the new $\lambda$, we repeat the second step to fourth step, until $\lambda$ converges. This method starts with $\lambda=0$, which corresponds to a scene that is a flat map(a map with the maximum possible entropy). The entropy of the scene decreases with each iteration as it converges on the scene with the maximum entropy that satisfies the constraints of the original map and the $PSF$. For more details, please refer to the Bonavito paper(\citealt{8bonavito1993maximum}).

\section{The simulation and verification}

The simulation includes three aspects, namely the simulation based on point source, the simulation based on an extended source and the simulation based on the Chang'E-2 microwave brightness temperature data, respectively. And two indices, Mean Squared Error(MSE) and Peak Signal to Noise Ratio(PSNR) are used to evaluate the quality of the deconvolution. A lower MSE and a higher PSNR indicate better deconvolution results. The definitions of the MSE and PSNR are given as below.

\begin{equation}\label{equ5}
   MSE = \frac{1}{{mn}}\sum\limits_{i = 0}^{m - 1} {\sum\limits_{j = 0}^{n - 1} {{{(Y(i,j) - X(i,j))}^2}} }
\end{equation}

\begin{equation}\label{equ6}
   PSNR = 10 \cdot {\log _{10}}(\frac{{MA{X^2}}}{{MSE}})
\end{equation}
where $X$ means the original map, $Y$ means the map to be evaluated and $MAX$ means the maximum value in $X$ data.

Here, a pattern function used as the fitting the antenna pattern is shown as bellow,
\begin{equation}\label{equ7}
 f(\theta ) = \sin c(\frac{{D * \sin (\theta )}}{\lambda })
\end{equation}
where $D$ is the diameter of the ground antenna, $\lambda $ is the observation wavelength and $\theta$ is the antenna pattern angle. The function is used in Simulation 3.1 and Simulation 3.2 .

\subsection{Simulation based on point sources}

\begin{figure}[htb]
   \centering
   \includegraphics[width=9.0cm, angle=0]{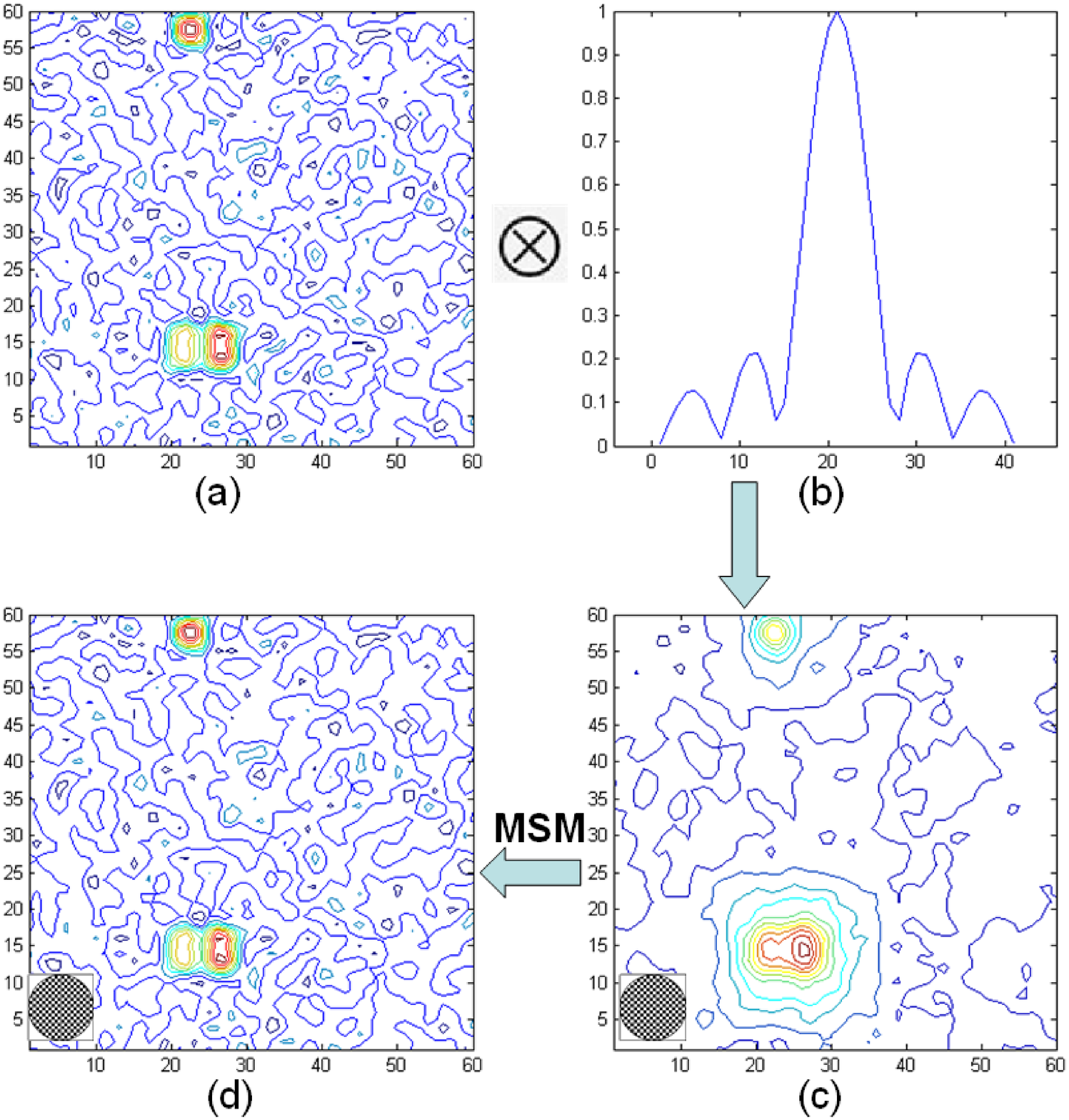}
   \caption{The flow figure of simulation based on the point sources (a) the original map (b)the antenna
pattern (c)the blurred map (d) the deconvolution map.}
   \label{fig2}
\end{figure}

The purpose of the simulation is to verify the validity of the method when dealing with a point source. The steps of simulation are as follows.

Firstly, in Figure 2a, the original data contains three point sources and values of the point sources are higher than the background. The second step is fitting an antenna pattern using Equation \ref{equ7}. Figure 2b shows a two-dimensional curve of the antenna pattern. The third step is convoluting the original data with the antenna pattern, then the blurred map is shown in Figure 2c. The fourth step is using the MEM to process the blurred map, and the result of deconvolution is shown in Figure 2d. In addition, the shadows on the lower left of Figure 2c and Figure 2d represent the projection of 3dB beam width on the original map and the deconvolution map, respectively.

Because of the impact of the beam width of antenna pattern, the two lower point sources in Figure 2c become nearly one point source. However, after the deconvolution of the MSE, the lower two point sources have been clearly distinguished in Figure 2d. Compared with the blurred map, the deconvolution map based the MEM recovers the details which can be found in the original data map. Moreover, we calculate the MSE and PSNR. The MSE and PSNR of the deconvolution map are 0.0113 and 58.3980 respectively, while the values are 0.6309 and 40.9298 for the blurred map. This result proves the process of deconvolution improves PSNR and reduces MSE.

\subsection{Simulation based on an extended source}

\begin{figure}[htb]
   \centering
   \includegraphics[width=14.0cm, angle=0]{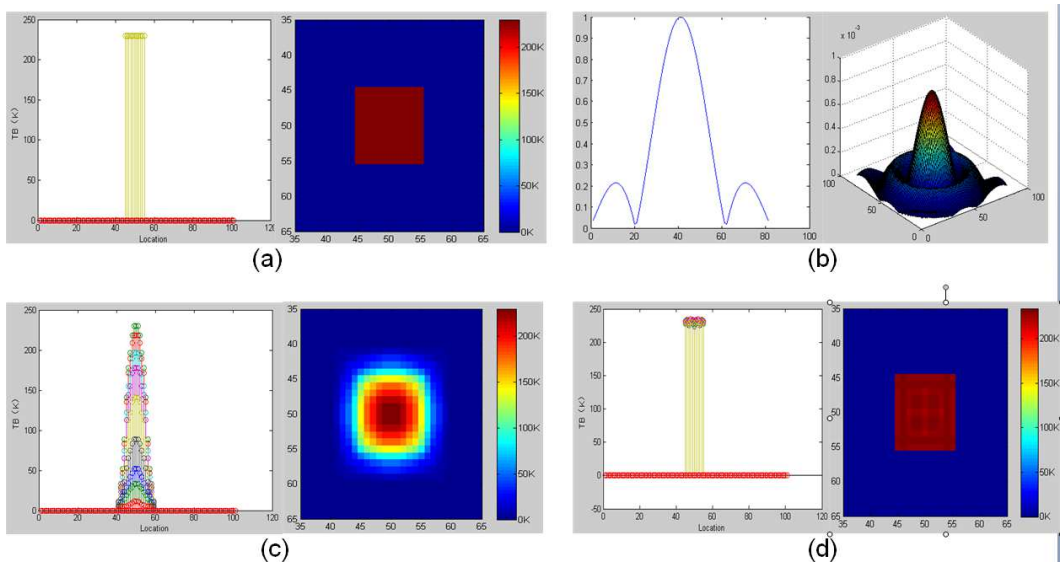}
   \caption{The simulation based on the ground antenna,(a) the original source (b) the antenna pattern
(c) the blurred map (d) the deconvolution map.}
   \label{fig3}
\end{figure}

\begin{figure}[htb]
   \centering
   \includegraphics[width=12.0cm, angle=0]{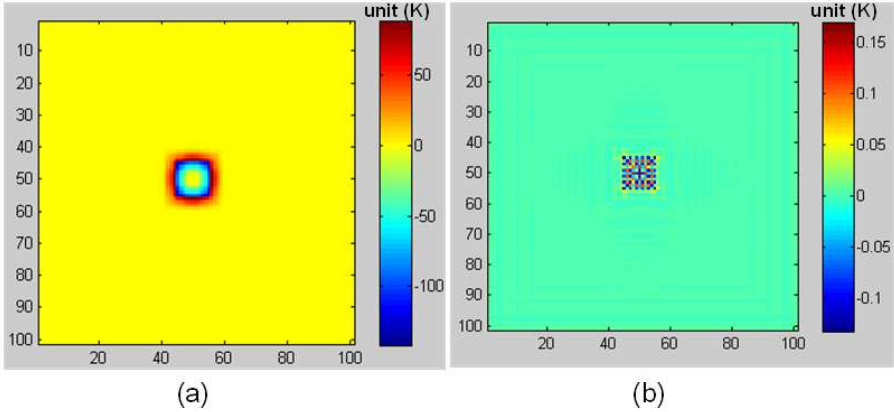}
   \caption{The residual map of the blurred map and the maximum entropy map (a) the blurred one
(b) the deconvolution one; the maps are abtained from the blurred map and the deconvolution map
minus the original map respectively.}
   \label{fig4}
\end{figure}

In this simulation, first, an original extended source is shown in Figure3a. The value in the middle area is 230k, while the value in other area is 0 k. And two forms, the discrete sequence image and the plane image are shown in order to describe the changes between different images clearly. Second, based on the Equation \ref{equ7}, the antenna pattern is shown in Figure 3b. Then, by convoluting the original data with the antenna pattern, the blurred image is shown in Figure 3c. Finally, using the MEM to process the blurred image, the result of deconvolution is shown in Figure 3d.

We calculate the MSE and PSNR. The MSE and PSNR of the blurred map are 0.6608 and 40.1589
respectively, while the values are 0.0466 and 51.6718 in the deconvolution map. Figure \ref{fig4} shows the residual
map of the blurred map and the deconvolution map. The value range of the residual map of the blurred map
is from -150 K to 100 K but the value range is about from -0.13 K to 0.17 K in deconvolution map. So when
processing the extended source data, the MEM produced a better result.

\subsection{Simulation based on Chang'E-2 microwave data}

During the ground calibration testing, we tested antenna pattern and gain of the 37GHz channel and then got four important parameters: (1) 3dB beam width of approximately $10^\circ$, (2) the main beam width of about $26^\circ$, (3) the attenuation of first side lobe is about 22dB, (4) the first side lobe beam width of approximately $6^\circ$. Based on Gaussian function, we fitted the antenna pattern function using the above four parameter(the contribution of main lobe is about 95.9\%(\citealt{10wang2009orbit}), so here we only considered the main lobe and first side lobe, which is shown in Equation \ref{equ7n}. The antenna pattern is shown in Figure \ref{fig5}.

Moreover, considering that bandwidth of 37GHz channel is about 500MHz, we  tested the antenna pattern and gain in three frequency points (36.47GHz, 37GHz, 37.53GHz)~during ground calibration testing. Taking 3dB beam width for instance, three frequency points(36.47GHz, 37GHz, 37.53GHz) corresponded to $10.07^\circ$, $10.17^\circ$ and $10.03^\circ$ respectively. The difference is less than $0.15^\circ$, which is about 0.25km corresponding to lunar surface. However, the spatial resolution of 37GHz channel is 13km, far larger than 0.25km. So in the paper we only use antenna pattern of 37GHz on behalf of the whole bandwidth.

\begin{equation}\label{equ7n}
f(\theta ) = \left\{ \begin{array}{l}
 - 30 \cdot {e^{ - {{(\frac{{\theta  + 13}}{{5.3}})}^2}}} - 30 \cdot {e^{ - {{(\frac{{\theta  - 13}}{{5.3}})}^2}}}\begin{array}{*{20}{c}}
{}&{\begin{array}{*{20}{c}}
{}&{\begin{array}{*{20}{c}}
{}&{}
\end{array}}&{\begin{array}{*{20}{c}}
{}&{}&{}&{}&{}
\end{array}}
\end{array}}&{ - {{13}^ \circ } \le \theta  \le {{13}^ \circ }}
\end{array}\\
 - 8 \cdot {e^{ - {{\left( {\frac{{\frac{{\left| \theta  \right| - 13}}{{0.23}}}}{{5.3}}} \right)}^2}}} - 8 \cdot {e^{ - {{\left( {\frac{{\frac{{\left| \theta  \right| - 13}}{{0.23}} - 26}}{{5.3}}} \right)}^2}}}{\rm{ - }}22\begin{array}{*{20}{c}}
{\begin{array}{*{20}{c}}
{}&{}
\end{array}}&{{{13}^ \circ } \le \left| \theta  \right| \le {{19}^ \circ }}
\end{array}
\end{array} \right.
\end{equation}

\begin{figure}[htb]
   \centering
   \includegraphics[width=14.0cm, angle=0]{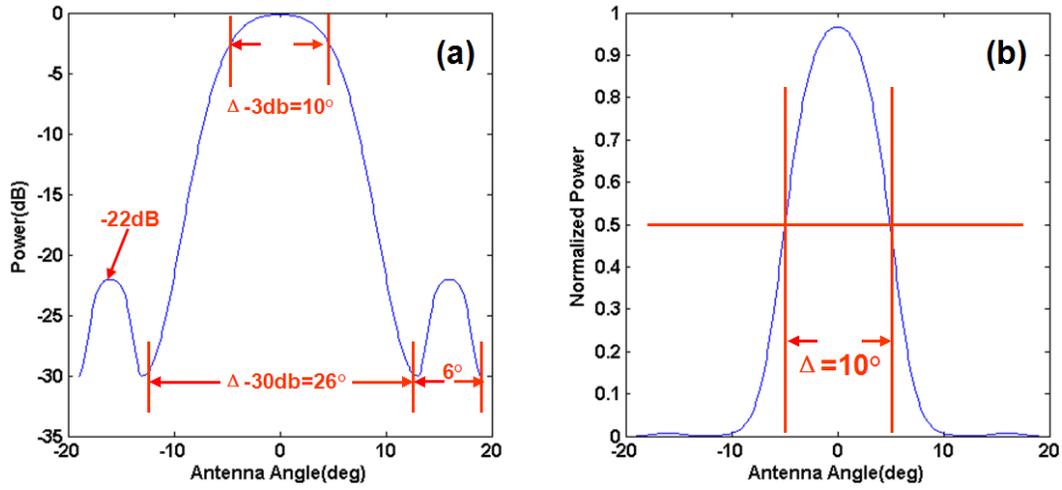}
   \caption{The antenna pattern of 37GHz, (a) Gain(dB) antenna pattern, (b) Normalized antenna pattern. According to Wang et al. 2009, the antenna pattern of 37GHz is figured out and used in the following deconvolution processing procedure.}
   \label{fig5}
\end{figure}

\begin{figure}[htb]
   \centering
   \includegraphics[width=14.0cm, angle=0]{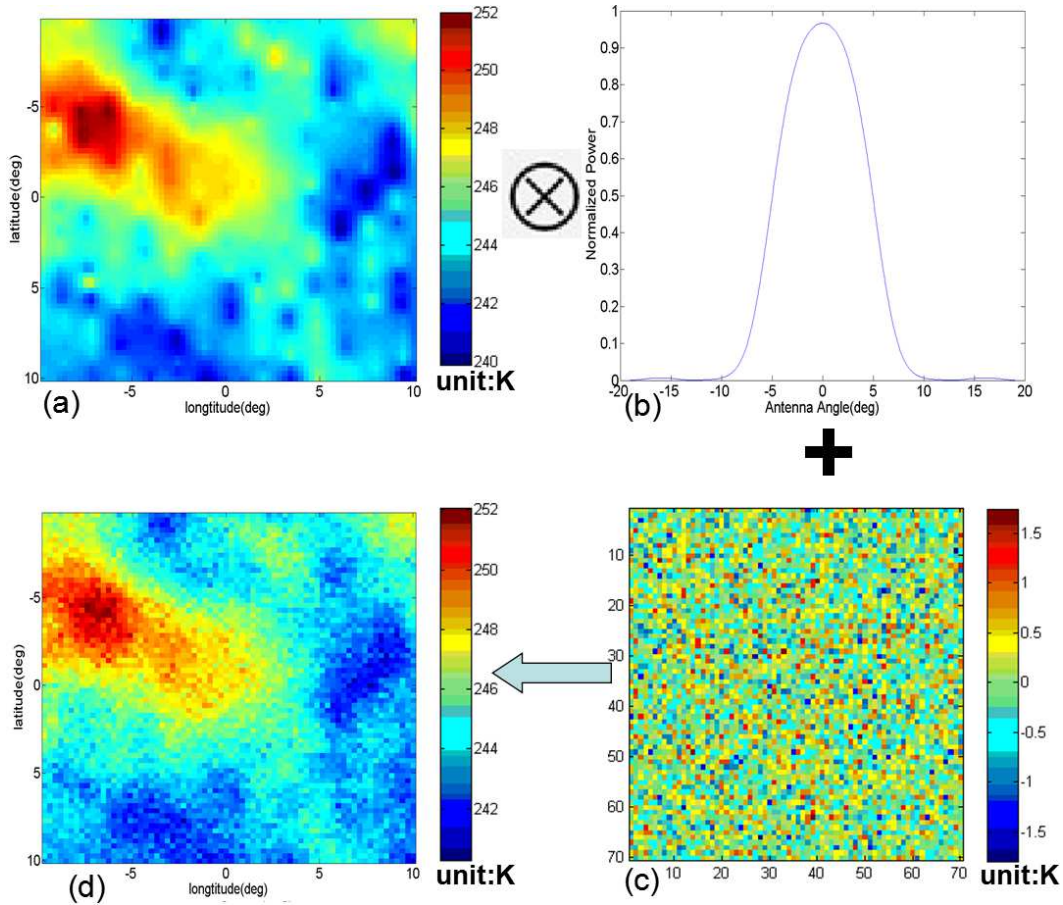}
   \caption{The result of simulation based on Chang¡¯E-2 microwave data (a) the original brightness
temperature map (b) the antenna pattern (c) the gaussian noise (d) the blurred map.}
   \label{fig6}
\end{figure}

\begin{figure}[htb]
   \centering
   \includegraphics[width=14.5cm, angle=0]{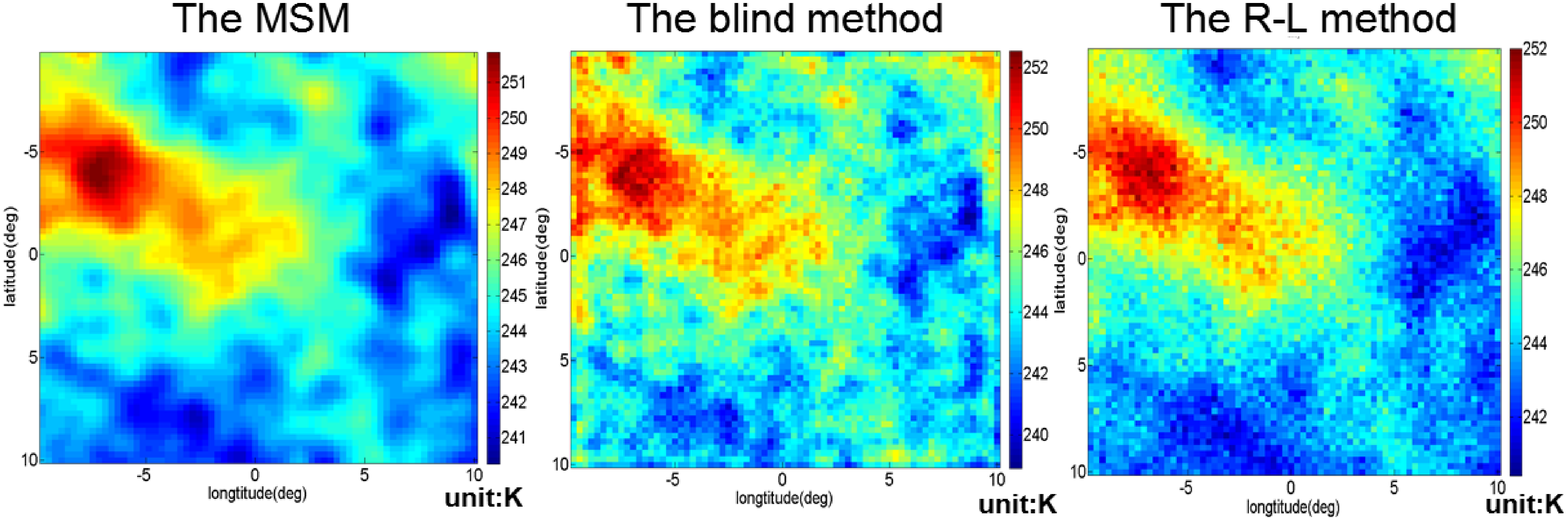}
   \caption{The deconvolution results from differernt methods.}
   \label{fig7}
\end{figure}

In this simulation, Chang'E-2 microwave data is used as the original data. The purpose of the simulation
is to verify the method used in this step can work well when processing the Chang'E-2 microwave data
with corresponding antenna pattern. Firstly, a region about ${20^ \circ } * {20^ \circ }$ chosen from Chang'E-2 microwave
brightness temperature map is shown in Figure 6a. As we choose the 37 GHz microwave data, so the
corresponding beam width of the antenna pattern is about $10^\circ$ (\citealt{10wang2009orbit})~and the antenna pattern
map is shown in Figure 6b. Given the sensitivity of the Chang'E-2 microwave sounder is approximately 0.5K (\citealt{10wang2009orbit}), a gaussian noise map is shown in Figure 6c, whose mean and standard deviation are
0 and 0.5 respectively. Figure 6d shows the blurred map, which is obtained by convoluting the original map
and the antenna pattern and then adding the noise.

Here, different methods such as the Richardson Lucy (R-L) deconvolution method and the blind decon-volution method (\citealt{11ayers1988iterative};\citealt{12biggs1997acceleration}) are chosen to process the same data, and we compare the processing results. The comparison is shown in Figure \ref{fig7}. And we also calculate the MSE and PSNR of the different method, which is shown is Table \ref{tab1}. By comparison, we find that the MEM shows a significant advantage both on restoring the details and restraining the noise.

\begin{table}[htb]
\bc
\begin{minipage}[]{100mm}
\caption[]{The PSNR and MSE of the different methods\label{tab1}}\end{minipage}
\setlength{\tabcolsep}{10pt}
\renewcommand{\arraystretch}{1.2}
 \begin{tabular}{ccccccccccccc}
 \hline
 ~~& the blurred & the MEM & the blind method & the P-L method\\
 \hline
 PSNR & 52.8063&57.8611&50.4692&52.7997\\
 MSE&0.3404&0.1039&1.2300&0.3333\\
 \hline
\end{tabular}
\ec
\tablecomments{0.65\textwidth}{The PSNR and MSE are calculate refer to Equation \ref{equ6} and \ref{equ5}.}
\end{table}

\clearpage
\subsection{Error Analysis}

Through the above three simulations, we found that the MEM is a good method to solve deconvolution problem. In this section, we discuss the radiometric error introduced by MEM.

\begin{figure}[htb]
   \centering
   \includegraphics[width=14.5cm, angle=0]{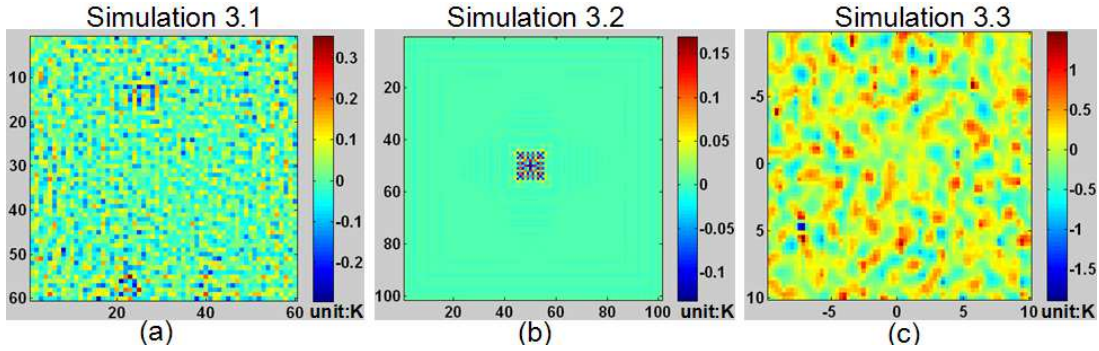}
   \caption{Error produced in the above three simulations, (a) simulation based on point sources  (b) simulation 3.2 based on extended source  (c) simulation based on real microwave data.}
   \label{fig8}
\end{figure}

Figure \ref{fig8} shows the error produced in above three simulations, which is got by the deconvolution map minus the original map respectively. We found that in simulation 1, the range of the error is from -0.35K to 0.4K with standard deviation $\sigma  = 0.0676K$; in simulation 2, the error is from -0.13K to 0.17K with $\sigma  = 0.0089K$, and in the simulation 3, because of  the Gaussian noise(-1.8K to1.8K) introduced, the error is about from -2K to 1.5K with $\sigma  = 0.3054K$. The errors introduced by MEM in the above three simulations are all less than the sensitivity of the microwave sounder which is 0.5K(\citealt{10wang2009orbit}).

Through the above simulations and error analysis, we can draw the conclusion that the MEM is an efficient method when dealing with the deconvolution problem and it can further be used to process the microwave brightness temperature data of the Chang'E-2 orbiter.

\clearpage
\section{Data Processing Based on Chang'E-2 Microwave Data}

In this section, the MEM is used to process the real microwave data. The data is the 37 GHz nighttime
microwave data acquired by Chang'E-2 orbiter, which is normalized with the Lunar local time. Refer to
the paper (\citealt{13zheng2012first}), we normalize the lunar hour angle to $0^\circ$, which implies the lunar time is the midday of the Moon.

There is also a problem worthy of notice, which is how to keep the scale of the microwave data consistent
with the antenna pattern. Figure \ref{fig9} describes the problem clearly. In this Figure, $h$ means the observation height of the antenna; $\theta$ means the angle of the antenna pattern; $r$ means the radius of the moon and $\alpha$ means the central angle of moon. In order to keep the consistent the projection of the $\theta$ and the $\alpha$ on the lunar surface, we get approximate equation shown in Equation \ref{equ8}. With further simplification, we get $\theta  \approx 17.4\alpha $, which means $1^ \circ$ on lunar surface corresponds $17.4^\circ$ in the antenna angle. When the microwave data value is taken every $0.2^ \circ$, the corresponding antenna pattern value is taken every $3.48^ \circ$. Thus, the correspondence of $0.2^ \circ$ in microwave data and $3.48^\circ$ in antenna pattern is adopted in this paper.

\begin{figure}[htb]
   \centering
   \includegraphics[width=11.0cm, angle=0]{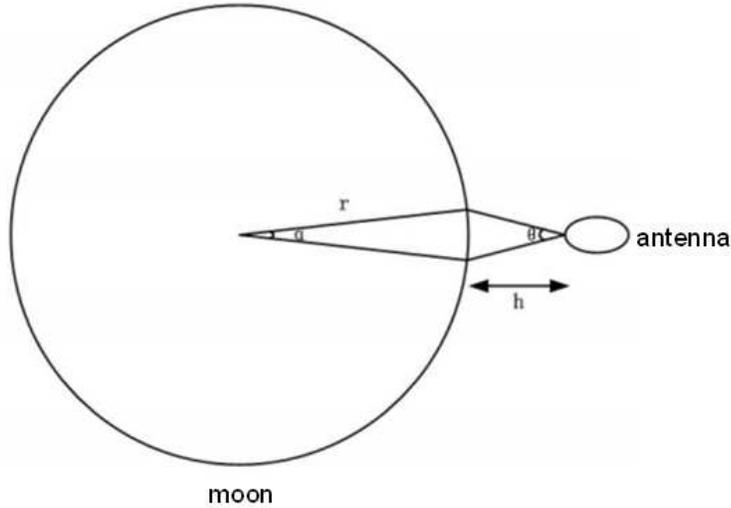}
   \caption{The sketch map of the antenna observation.}
   \label{fig9}
\end{figure}

\begin{equation}\label{equ8}
    \frac{\alpha }{{180}} \times \pi  \times r \approx \frac{\theta }{{180}} \times \pi  \times h
\end{equation}

In addition, the Digital Elevation Model(DEM) data and Charge Coupled Device(CCD) data from
Chang'E-2 orbiter are used in this section. The laser altimeter and the CCD camera are also the main
payloads on the Chang'E-2 orbiter and the spatial resolution of the two payloads is higher than the microwave sounder. So the data are used to verify the validity of the new added details in the deconvolution map. The comparison of the original distribution map and the deconvolution distribution map based on
MEM is shown in Figure \ref{fig10}.

Furthermore, three regions are also chosen to show the differences between the original map and the
deconvolution map. The range of first region is from $14.3^ \circ$W to $11.27^ \circ$E and $48.2^ \circ$S to $22.4^ \circ$S. The second region is about ${10^ \circ } * {10^ \circ }$ with centering the Apollo 16 landing point, while the third region is centered the
Luna 24 landing point. Comparisons of three regions are shown in Figure \ref{fig11},\ref{fig12} and \ref{fig13} respectively.

\begin{figure}[htb]
   \centering
   \includegraphics[width=13.0cm, angle=0]{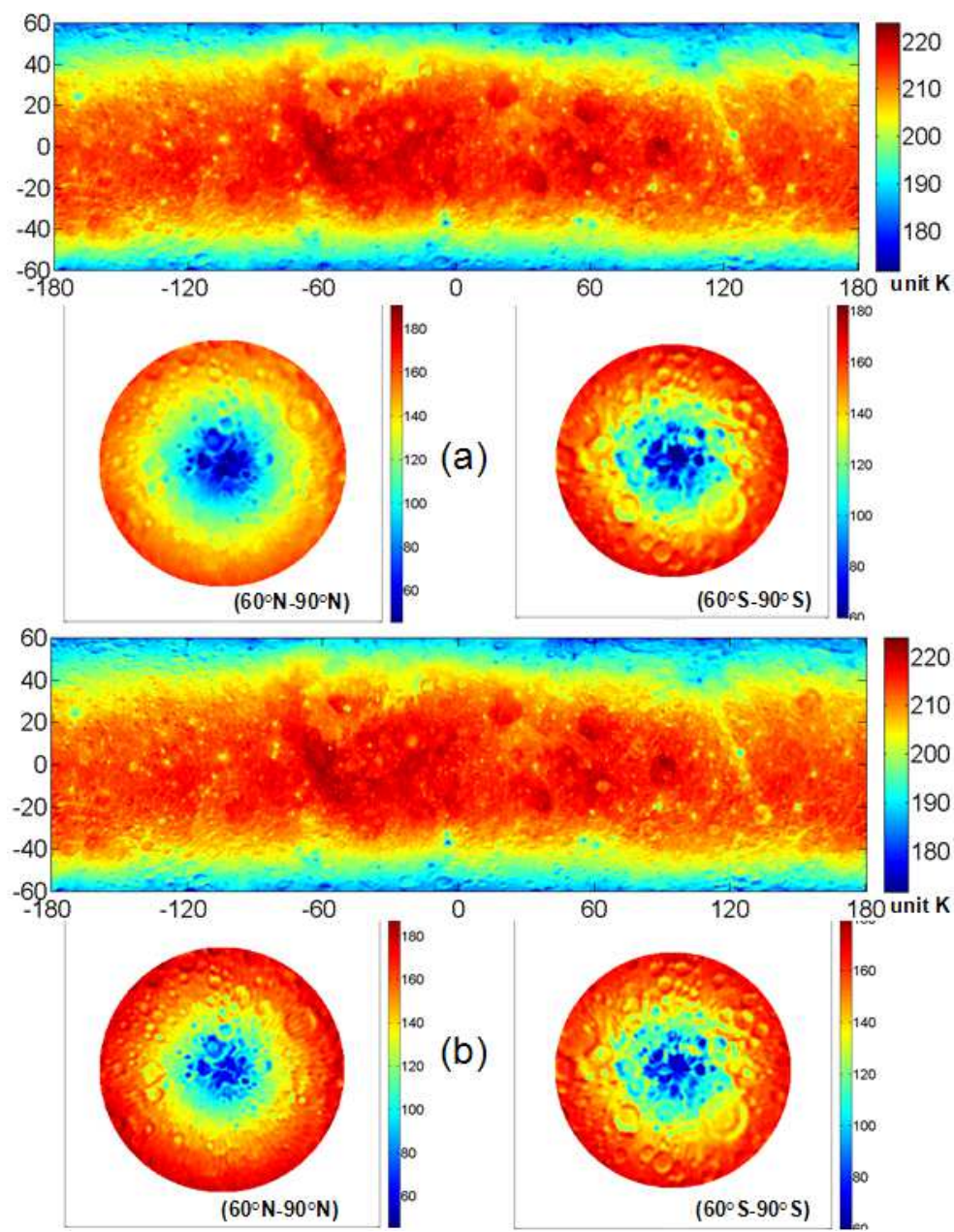}
   \caption{The comparison of the original map and the deconvolution map of the whole lunar surface,(a)the original map, (b) the deconvolution map. The original data came from the 37GHz
nighttime microwave data.}
   \label{fig10}
\end{figure}

\begin{figure}[htb]
   \centering
   \includegraphics[width=13.0cm, angle=0]{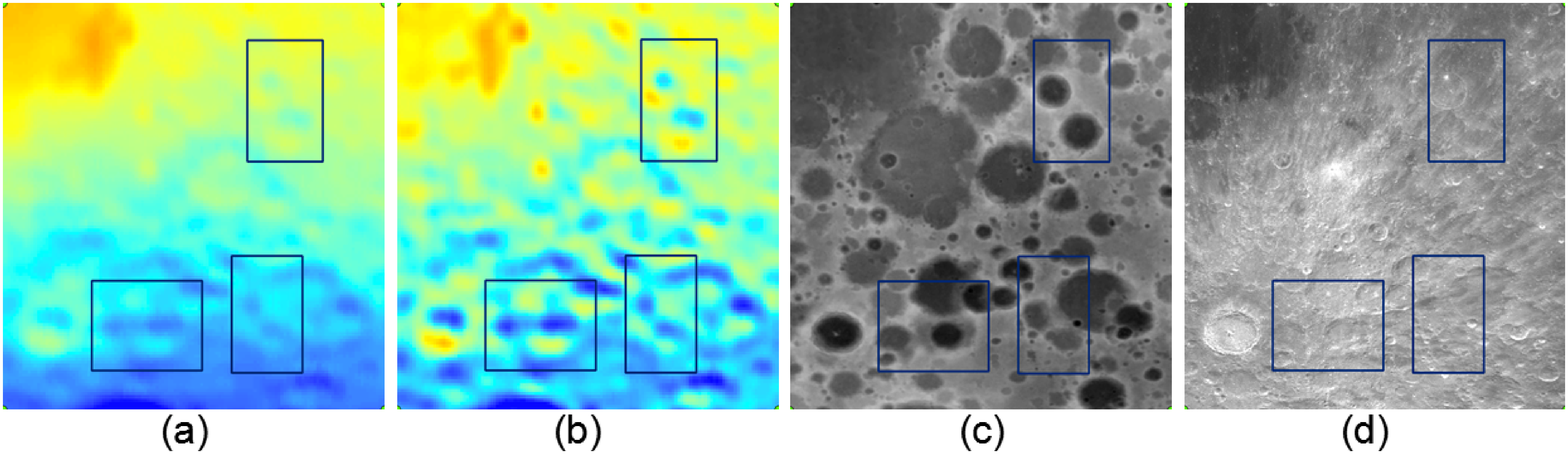}
   \caption{The comparison result map of the region with the longitude range from $14. 3^\circ$W
to $11.27^\circ$E, the latitude range from $48.2^\circ$S to $22.4^\circ$S. (a) the original brightness temperature map, (b) the deconvolution brightness temperature map, (c) the DEM map, (d) the CCD map.}
   \label{fig11}
\end{figure}

\begin{figure}[htb]
   \centering
   \includegraphics[width=13.0cm, angle=0]{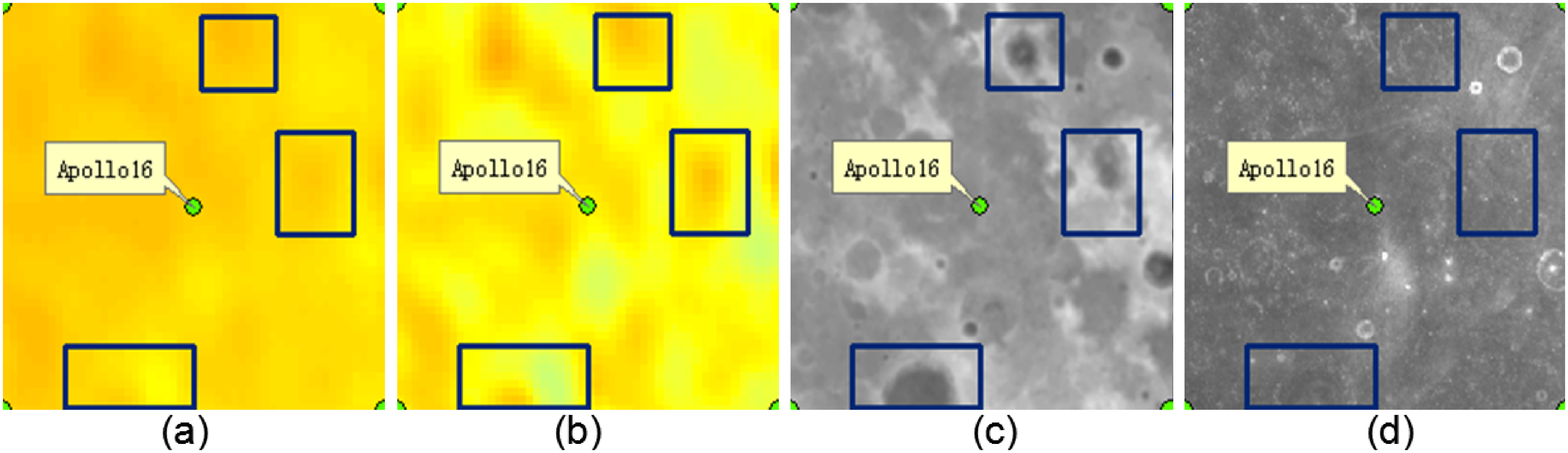}
   \caption{The comparison result map of the region with the longitude range from $57.2^\circ$E to $67.2^\circ$E, the latitude range from $7.3^\circ$N to $17.3^\circ$N. (a) the original brightness temperature map, (b)the deconvolution brightness temperature map, (c) the DEM map, (d) the CCD map.}
   \label{fig12}
\end{figure}

\begin{figure}[htb]
   \centering
   \includegraphics[width=13.0cm, angle=0]{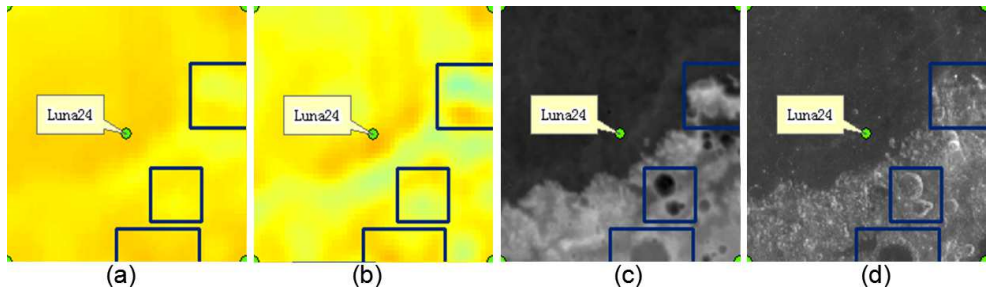}
   \caption{The comparison result map of the region with the longitude range from $10.5^\circ$E to $20.5^\circ$E, the latitude range from $14^\circ$S to $4^\circ$S. (a) the original brightness temperature map, (b)the deconvolution brightness temperature map, (c) the DEM map, (d) the CCD map.}
   \label{fig13}
\end{figure}

Figure \ref{fig10} indicates that the deconvolution brightness temperature distribution map has higher resolution than the original map. In order to clearly compare the deconvolution map and the original map, we chose three regions. From Figure \ref{fig11}, \ref{fig12} and \ref{fig13}, we find that there are some new added details which have bounded by blue lines in the deconvolution map. Furthermore, in the position of the domains, we find similar features in both the DEM and the CCD map.

In deconvolution maps, under the conditions of only considering the influence of the antenna pattern, the brightness temperature error introduced by MEM is less than 0.5K.

\clearpage

\section{Conclusion}

In this paper, the MEM is first used to deconvolute the Chang'E-2 microwave data and the results are
satisfactory. Compared to the original brightness temperature map, the map processed by MEM has more
details and higher resolution. Only considering the influence of the antenna pattern, the new data obtained by MEM processing is very useful for further research and the brightness temperature error introduced by MEM is less than 0.5K. We conclude the following.

Given the impact of the antenna pattern, it is necessary to find a suitable deconvolution method used
to process Chang'E-2 microwave data. And by applying the simulations above, we found on the one hand
the impact of the antenna pattern could blur the original map heavily, but on the other hand MEM was effective in increasing PSNR and reducing MSE,  for both extended sources and point sources, eventually restoring the original distribution information. Furthermore, in comparison to other methods MEM shows significant advantages, both for restoring detail and reducing noise.

After comparison of the whole lunar brightness temperature distribution map and several local regions,
we found the deconvolution maps got more details and higher resolution than the original maps. So the
more precise and higher resolution brightness temperature has been achieved.

By comparison between the brightness temperature map and the topographic map, we found the new
added details also have a correspondence in the CCD map and DEM map. This, for one thing, verifies the
validity of the new added details, for another reveals a relativity between brightness temperature and
lunar terrain and geological formation. Thus, the precise brightness temperature data processed by MEM is
very useful to study the lunar surface and subsurface structure for further research.

\normalem
\begin{acknowledgements}
The work is supported by the National Natural Science Foundation of China (Grant
Nos.11173038 and 11203046). We thank the Lunar and Deep Space Exploration Department, National Astronomical Observatories, Chinese Academy of Sciences(NAOC) for providing the microwave data, CCD and DEM map.
\end{acknowledgements}

\bibliographystyle{raa}
\bibliography{bibtex}

\end{document}